\documentclass[12pt]{article}

\usepackage{amsmath,amssymb}
\usepackage[dvips]{graphicx,color}
\usepackage[round]{natbib}
\usepackage{mathrsfs}
\usepackage{bm, xcolor}
\usepackage{mathrsfs}
\usepackage{verbatim}
\usepackage{threeparttable}
\usepackage{color}
\usepackage[utf8]{inputenc}

\usepackage{graphicx}
\usepackage{subfigure}
\usepackage{float}
\usepackage{listings}
\usepackage{amssymb}
\usepackage{amsthm}
\usepackage{multirow}

\usepackage{caption}
\usepackage{filecontents}

\renewcommand{\baselinestretch}{1.3}

\makeatletter
\def\singlespace{\def\baselinestretch{1}\@normalsize}

\newtheorem{lemma}{Lemma}
\newtheorem{theorem}{Theorem}

\@addtoreset{equation}{section}

\renewcommand{\hat}{\widehat}

\def\singlespace{\def\baselinestretch{1}\@normalsize}

\makeatother

\newdimen\biblioindent    \biblioindent=30pt

 at 7truept

\def\beq{\begin{equation}}
\def\eeq{\end{equation}}
\def\beqn{\begin{eqnarray}}
\def\eeqn{\end{eqnarray}}
\def\beqnn{\begin{eqnarray*}}
\def\eeqnn{\end{eqnarray*}}

\begin{document}
\title{Review on Determining the Number of Communities in Network Data}
\author{Zhengyuan Du and Jason Cui}
\maketitle
\begin{abstract}
This paper reviews statistical methods for hypothesis testing and clustering in network models. We analyze the method by Bickel et al. (2016) for deriving the asymptotic null distribution of the largest eigenvalue, noting its slow convergence and the need for bootstrap corrections. The SCORE method by Jin et al. (2015) and the NCV method by Chen et al. (2018) are evaluated for their efficacy in clustering within Degree-Corrected Block Models, with NCV facing challenges due to its time-intensive nature. We suggest exploring eigenvector entry distributions as a potential efficiency improvement.
\end{abstract}
\section{Introduction}
Network-structured data and network analysis are garnering increasing attention across various fields. Research in this area often focuses on understanding the structure of network data, with significant implications for social sciences, biology, and statistics. The practical applications of this research profoundly impact our daily lives in multiple ways. For example, search engines utilize discoveries and tools from this field to analyze the relationships among various keywords. See \citet{kolaczyk2014statistical}, \citet{hevey2018network}, \citet{sun2013mining}, and \citet{berkowitz2013introduction} for further reading.

Community detection is of major interest of network analysis. Given an $n$-node (undirected) graph $(\mathcal{N}, E)$, where $\mathcal{N}=\{1,2,...n\}$ is the set of nodes and $E$ is the set of edges.  We assume that $\mathcal{N}$ could be partitioned into $K$ disjoint subsets or "communities". The community structure is represented by a vector $g = (g_1,..., g_n)$ with $g_i \in \{1,..., K\}$ being the community that node $i$ belongs to. Nodes tend to have more common characteristics in the same communities. Various algorithms have been proposed to partition these nodes. However, most of them must fixed $K$ as a priori. How to determine $K$ priori is still an open problem.

 \citet{bickel2016hypothesis} proposed a testing statistics based on the limiting distribution of the principal eigenvalue of the suitably centred and scaled adjacency matrix. However they only have theoretical results for testing null hypothesis of $H_0: K=1$. For testing $H_0: K=j$ when $j>1$, they need to iteratively split the network into small sub-networks. It seems that their testing statistics does not work well on the later one.  Compared to stochastic block model  (SBM), degree corrected block model  (DCBM) proposed by Karrer and Newman  (2011) is more flexible. \citet{jin2015fast} proposed a so called spectral clustering on ratios-of-eigenvectors (SCORE) method to detect communities for DCBM. The idea is by clustering on entry-wise ratio of eigenvectors of the adjacency matrix. Taking the ratio can eliminate the effect of degree heterogeneity. Further \citet{ji2016coauthorship} apply SCORE to a practical problem on detecting potential communities in coauthorship and citation networks for statisticians. However,  there are few works on the hypothesis testing of $K$ targetedly for DCBM. Instead of performing hypothesis testing, \citet{chen2018network} proposed a specially designed cross-validation method to figure out $K$.  In the following, we are going to give a brief review and duplicate the numerical results of some of the aforementioned papers.
\section{\citet{bickel2016hypothesis}}
In an SBM with $n$ nodes and $K$ communities, Let $A$ be the adjacency matrix. and $g = (g_1, ... ,g_n)$ with $g_i \in \{1, ... ,K\}$ being the community that node $i$ belongs to. Given the membership vector $g$, each edge $A_{ij}  (i < j)$ is an independent Bernoulli variable satisfying
\begin{equation}
P\left(A_{i j}=1\right)=1-P\left(A_{i j}=0\right)=B_{g_{i} g_{j}},
\end{equation}
where $B$ is a $K \times K$ symmetric matrix representing the community-wise edge probabilities.

\citet{bickel2016hypothesis} proposed a statistics on testing problem with null hypothesis $H_0: K=1$ based on some  properties of Erd\H{o}s-Ren\'yi graph. In their paper, they assume that the number of clusters $K$ and the edge probabilities are constant, whereas the number of nodes $n$ is growing to $\infty$. Thus the average degree is growing linearly with $n$. In addition \citet{cragg1993testing}, \citet{cragg1997inferring}, \citet{cui2024hypothesis}, \citet{cui2024double} \citet{du2025short} and \citet{cui2023enhancing} discussed about rank inference of a matrix, which can be a generalization work of \citet{bickel2016hypothesis}.

Noticing that Erd\H{o}s-Ren\'yi graph can be viewed as a special case of SBM when there is only one community and one community-wise edge probability $p:=B_{11}$. We can estimate it within $O_P(1/n)$ error by computing the proportion of pairs of nodes that forms an edge, denoted by $\hat{p}$.

Let $\hat{P}:=n \hat{p} \mathbf{e} \mathbf{e}^{\mathrm{T}}-\hat{p} I$, which is an estimate for the mean of adjacency matrix. \citet{bickel2016hypothesis} proved the following theorem:

\begin{theorem}
(\citet{bickel2016hypothesis}) Let
\begin{equation}
\tilde{A}^{\prime}:=\frac{A-\hat{P}}{\sqrt{\{(n-1) \hat{p}(1-\hat{p})\}}}
\end{equation}
We have the following asymptotic distribution of our test statistic $\theta$:
\begin{equation}
\theta:=n^{2 / 3}\left\{\lambda_{1}\left(\tilde{A}^{\prime}\right)-2\right\} \stackrel{\mathrm{d}}{\rightarrow} \mathrm{TW}_{1}
\end{equation}
where $\mathrm{TW}_1$ denotes the Tracy–Widom law with index 1. This is also the limiting law of the largest eigenvalue of GOEs.
\end{theorem}

\emph{Example 1.} 
Here we performed a simulation study to see the size performance of the testing statistics. We construct a Erd\H{o}s-Ren\'yi graph with $n= 500$. $p$ ranges from 0.1 to 0.9, increasing with 0.1 in each step.
\begin{figure}
\begin{center}
\includegraphics[width=5in]{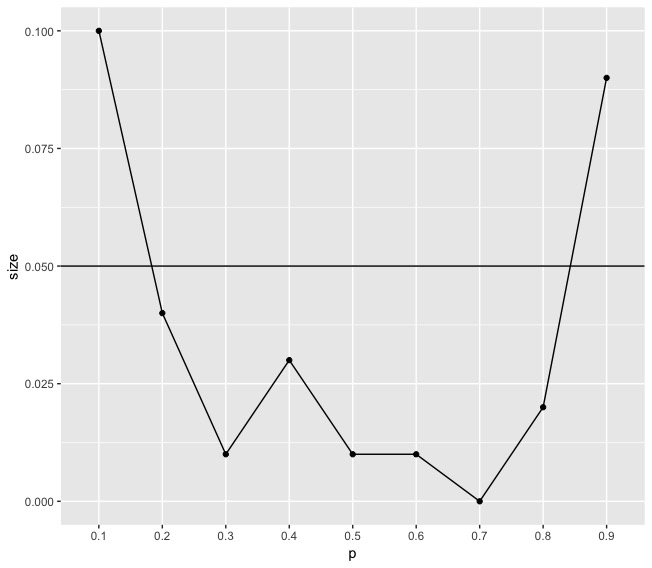}\caption{Empirical size of $\theta$ under 500 replications at significant level 0.05. The  horizontal  solid  line  represents significant level 0.05. }
\end{center}
\end{figure}

We can see from the simulation study that they cannot control the size. The author proposed a small sample correction on their testing statistics. That is due to the low convergence rate of the eigenvalues, whereas the largest eigenvalues of GOE matrices converge to the Tracy–Widom distribution quite quickly, those of adjacency matrices do not. Based on that, they proposed a small sample correction which  suggests that computing the $p$-value by using the empirical distribution of $\lambda_1$ generated by using a parametric bootstrap step. The main idea is to compute the mean and the variance of the distributions from a few simulations, and then shift and scale the test statistic to match the first two moments of the limiting TW$_1$ law. Table \ref{tab1} shows the whole algorithm for perfoming the hypothesis proposed in \citet{bickel2016hypothesis}.

\begin{table}\caption{Algorithm for hypothesis test in \citet{bickel2016hypothesis} with correction}\label{tab1}
\begin{center}
\fbox{\shortstack[l]{
\emph{Step. 1: } $\hat{p}=\Sigma_{i j} A_{i j} /\{n(n-1)\}$\\
\emph{Step. 2:} $\left.\theta \leftarrow n^{2 / 3}\left(\lambda_{1}[(A-\hat{P}) / \sqrt{\{}(n-1) \hat{p}(1-\hat{p})\}\right]-2\right)$ \\
\emph{Step. 3:}  $\begin{array}{l}{\mu_{\mathrm{TW}} \leftarrow E_{\mathrm{TW}_{1}}[X]} \\ {\sigma_{\mathrm{TW}} \leftarrow \sqrt{\operatorname{var}_{\mathrm{TW}}(X)}}\end{array}$\\
$\text { for } i=1, \ldots .50 \text { do }$\\
\emph{Step 4:} $\begin{array}{l}{A_{i} \leftarrow \text { Erdós-Renyi }(n, \bar{p})} \\ {\theta_i \leftarrow n^{2 / 3}\left[\lambda_{1}(A-\hat{P}) / \sqrt{/}\{(n-1) \hat{p}(1-\hat{p})\}-2\right]}\end{array}$\\
\emph{Step 4:}
$\begin{array}{l}{\hat{\mu}_{n, \hat{p}} \leftarrow \operatorname{mean}\left(\left\{\theta_{i}\right\}\right)} \\ {\hat{\sigma}_{n, \hat{p}} \leftarrow \text { standard deviation }\left(\left\{\theta_{i}\right\}\right)}\end{array}$\\
\emph{Step 5:}
$\theta^{\prime} \leftarrow \mu_{\mathrm{TW}}+\left\{\left(\theta-\hat{\mu}_{n, \hat{p}}\right) / \hat{\sigma}_{n, \hat{p}}\right\} \sigma_{\mathrm{TW}}$\\
\emph{Step 6:}
$\text { pval } \leftarrow P_{\mathrm{TW}_{1}}\left(X>\theta^{\prime}\right)$
}}
\end{center}
\end{table}

\section{Spectral Clustering Method for DCBM}
In this section we want to look at a spectral clustering method for and DCBM, the SCORE method proposed by \citet{jin2015fast}.

In a DCBM, given membership vector $g$ and community-wise connectivity matrix $B$, Let $\psi$ and $\Psi$ be the $n\times 1$ vector and the $n\times n$ diagonal matrix defined as follows: $$\psi=(\psi_1, \psi_2, \ldots, \psi_n)^{\prime}, \quad \Psi(i, i)=\psi(i), \quad 1 \leq i \leq n.$$
The presence of an edge between nodes $i$ and $j$ is represented by a Bernoulli random variable $A_{ij}$ with
$$P\left(A_{i j}=1\right)=1-P\left(A_{i j}=0\right)=\psi_{i} \psi_{j} B_{g_{i} g_{j}}$$
The main difference between DCBM and SBM is the 
appearance of the degree heterogeneity parameter $\psi_i>0$. $\psi_i$ represents the individual activeness of node $i$. The idea of that is that some individuals in one group could be more outgoing than others, therefore edge probabilities are different for different individuals.

 While the traditional spectral clustering method of SBM is simply by performing existing clustering methods on the eigenvectors of adjacency matrix, detecting communities with the DCBM is not an easy problem, where the main challenge lies in the degree heterogeneity.
 
The main observation of SCORE is that the effect of degree heterogeneity is largely an- cillary, and can be effectively removed by taking entry-wise ratios between eigenvectors. 
Let's firstly consider the simple case when there are only two groups. Specifically, let
$$P(A_{ij}=1)=\psi_i \psi_j\left\{\begin{array}{ll}{a,} & {g_i= g_j=1,} \\ {c,} & {g_i= g_j=2,} \\ {b,} & {\text { otherwise }}\end{array}\right.$$
and denote $\Omega=E(A)$
Theorem \ref{thm2} is

\begin{lemma}\label{thm2}
(\citet{jin2015fast}) If $ac\neq b^2$, then $\Omega$ has two simple nonzero eigenvalues
$$\frac{1}{2}\|\psi\|^{2}\left(a d_{1}^{2}+c d_{2}^{2} \pm \sqrt{\left(a d_{1}^{2}-c d_{2}^{2}\right)^{2}+4 b^{2} d_{1}^{2} d_{2}^{2}}\right)$$
and the associated eigenvectors $\eta_1$ and $\eta_2$ (with possible nonunit norms) are

$$\Psi\left(b d_{2}^{2} \cdot \mathbf{1}_{1}+\frac{1}{2}\left[c d_{2}^{2}-a d_{1}^{2} \pm \sqrt{\left(a d_{1}^{2}-c d_{2}^{2}\right)^{2}+4 b^{2} d_{1}^{2} d_{2}^{2}}\right] \cdot \mathbf{1}_{2}\right)$$
\end{lemma}

The key observation is that if we let $r$ be the $n \times 1$ vector of the coordinate- wise ratios between $\eta_1$ and $\eta_2$. $r_i=\dfrac{\eta_{2i} /\left\|\eta_{2}\right\|}{\eta_{1i} /\left\|\eta_{1}\right\|}, \quad 1 \leq i \leq n$. Then $r$ is independent from the degree heterogeneity parameter $\psi$.

Similar phenomenon also exists when there are $K>2$ groups.

Based on that, the algorithm of clustering on DCBM is proposed as following:

\begin{itemize}
\item Let $\hat{\eta}_1$ , $\hat{\eta}_2$ ,..., $\hat{\eta}_K$ be $K$ unit-norm eigenvectors of $A$ associated with the largest K eigenvalues (in magnitude), respectively.
    
\item Let $\hat{R}$ be the $n\times K-1$ vector of coordinate-wise ratios: $\hat{R}(i, k)=\hat{\eta}_{k+1,i} / \hat{\eta}_{1,i}, \quad 1 \leq k \leq K-1,1 \leq i \leq n$
\item Clustering the labels by applying the k-means method to the vector
$\hat{R}$, assuming there are $\leq K$ communities in total.
\end{itemize}

\begin{figure}
\begin{minipage}[t]{0.5\linewidth}
\includegraphics[width=3in]{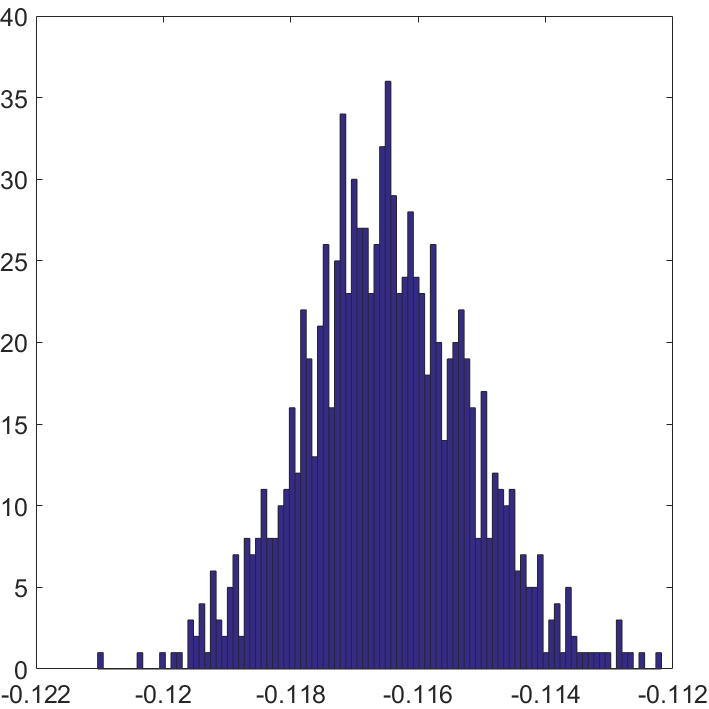}
\end{minipage}%
\begin{minipage}[t]{0.5\linewidth}
\includegraphics[width=3in]{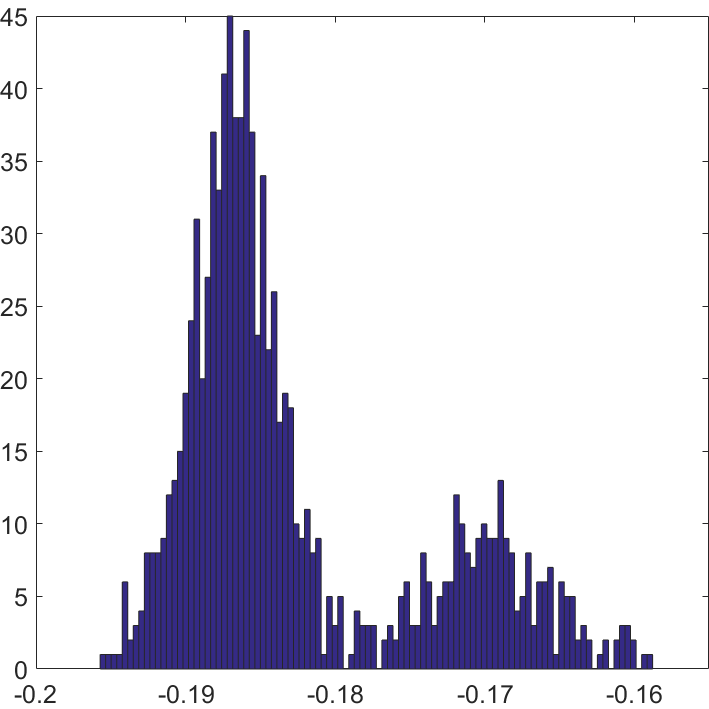}
\end{minipage}\caption{\emph{The vector $\hat{R}$, Left:the case when $K=1$. Right: the case when $K=2$}}\label{fig2}
\end{figure}

\emph{Example 2.} Here I performed one simulation study to show the intuition of SCORE.
We take $K=1$ and $K=2$ respectively to construct networks. For both we take $n=1000$ and generate $\bm{\psi}$ by $\text{log} (\psi_i) \overset{i.i.d.}{\sim} N (0,0.2), 1 \leq i \leq n,$ then normalize $\psi$ by $\psi=0.9\cdot\psi/\psi_{\text{max}}.$  For $K = 2$,  250 nodes are in different cluster with others.  $B_{11}=B_{22}=1,B_{12}=0.5.$ The coordinate-wise plot  is reported in Figure \ref{fig2}. Here when $K=1$, the $\hat{R}$ is given by computing the coordinate ratio of $\eta_1$ and $\eta_2$. The patterns between $K=1$ and $K=2$ have significantly difference. If $K=1$, each coordinate of $\hat{R}$ will gather around one constant as we showed before, while it will on two hierarchies rather than one when $K=2$.

\section{\citet{chen2018network}}
\citet{chen2018network} proposed a network cross-validation (NCV) approach
to determine the number of communities for both SBM and DCBM. It can automatically determine the number of communities based on a block-wise node-pair splitting technique.
The main idea is that firstly, we randomly split the nodes into $V$ equal-sized subsets $\left\{\tilde{\mathcal{N}}_{v}: 1 \leq v \leq V\right\}$, and then split the adjacency matrix correspondingly into $V \times V$ equal sized blocks. 
\begin{equation}\label{eq1}
    A=\left(\tilde{A}^{(u v)}: 1 \leq u, v \leq V\right)
\end{equation}
where $\tilde{A}^{(u v)}$ is the submatrix of $A$ with rows in $\tilde{N}_{u}$ and
columns in $\tilde{\mathcal{N}}_{v}$.
All the blocks are splitted into fitting sets and validation sets.  We can estimate model parameters $\left(\hat{g}^{(v)}, \hat{B}^{(v)}\right)$ using the rectangular submatrix as fitting set obtained by removing the rows of $A$ in subset $\tilde { N } _ { v }$.

\begin{equation}
\tilde{A}^{(-v)}=\left(\tilde{A}^{(r s)}: r \neq v, 1 \leq r, s \leq V\right).
\end{equation}

Then $\tilde{A}^{(v v)}$ is treated as validation sets and used to calculate the predictive loss. At last, $K$ is selected when the loss is minimized.
In general, cross-validation methods are insensitive to the number of folds. The same intuition empirically holds true for the proposed NCV method. Here the author suggest to use $V=3$.

Take $V=2$ for example. We can rearrange adjacency matrix in a collapsed $2 \times 2$ block form 
$$A=\left(\begin{array}{ll}{A^{(11)}} & {A^{(12)}} \\ {A^{(21)}} & {A^{(22)}}\end{array}\right)$$
Such a splitting puts node pairs in $A^{(11)}$ and $A^{(12)}$ as the fitting sample and those in $A^{(22)}$ as the validating sample. 
The advantages of that kind of splitting are from three aspects. Firstly, the fitting set carries full
information about the network model parameters. We can consistently estimate the
membership of all the nodes as well as the community-wise edge probability matrix, using only data in the fitting set. Secondly,  given the community membership, the data in the fitting set and in the testing set are independent. Thirdly, it is different from cross validation methods for network data based on a node splitting technique. In the node splitting method, the nodes are usually split into a fitting set and a testing set. Therefore, one typically assumes that the node memberships are generated independently with prior probability $\pi=(\pi_1,\pi_2,\cdots,\pi_K)$. However, it  has several drawbacks. The main drawback is that calculating the full likelihood in terms of the prior probability in the presence of a missing membership vector $g$ is computationally demanding. 

Algorithm details of their NCV method is shown as following:
\begin{itemize}
    \item \emph{Step 1:} Block-wise node-pair splitting as shown in $(\ref{eq1})$
    \item \emph {Step 2:} Estimating model parameters from the fitting set.
    
    Take SBM for example.
    $$\begin{array}{l}{\text { 1. Let } \widehat{U} \text { be the } n \times d \text { matrix consisting of the top } d \text { right singular vectors of } A^{(1)}} \\ {\text { 2. Output } \widehat{g} \text { by applying the } k \text { -means clustering with } \widetilde{K} \text { clusters to the rows of\ } \widehat{U}. }\end{array}$$
    Once $\hat{g}$ is obtained, let $\mathcal{N}_{j, k}$ be the nodes in  $j$ with estimated membership $k$,and $n_{i, k}=\left|\mathcal{N}_{i, k}\right|(j=1,2,1 \leq k \leq \tilde{K})$.Wecan estimate $B$ using a simple plug-in estimator:
   
   $$\hat{B}_{k, k^{\prime}}=\left\{\begin{array}{ll}{\frac{\sum_{i \in \mathcal{N}_{1, k}, j \in \mathcal{N}_{1, k^{\prime}} \cup N_{2, k^{\prime}}} A_{i j}}{n_{1, k}\left(n_{1, k^{+}}+n_{2, k^{\prime}}\right)},} & {k \neq k^{\prime}} \\ {\frac{\sum_{i, j \in \mathcal{N}_{1, k}, i<j} A_{i j}+\sum_{i \in \mathcal{N}_{1, k}, j \in \mathcal{N}_{2, k} A_{i j}}}{\left(n_{1, k}-1\right) n_{1, k} / 2+n_{1, k} n_{2, k}},} & {k=k^{\prime}}\end{array}\right.$$
    \item \emph{Step 3:} Validation using the testing set.
    $\text {Consider the validated predictive loss } \widehat{L}(A, \widetilde{K})=\sum_{i, j \in \mathcal{N}_{2}, i \neq j} \ell\left(A_{i j}, \widehat{P}_{i j}\right)$, with $\text { negative log-likelihood } \ell(x, p)=-x \log p-(1-x) \log (1-p)$ or squared error $\ell(x, p)=(x-p)^{2}$.
\end{itemize}

Algorithm of dealing with DCBM is similar. The only difference is that in step 2, we need to estimate the parameters $(g,B, \psi)$ under the framework of DCBM.

\begin{figure}[h]
\begin{minipage}[t]{0.5\linewidth}
\includegraphics[width=3in]{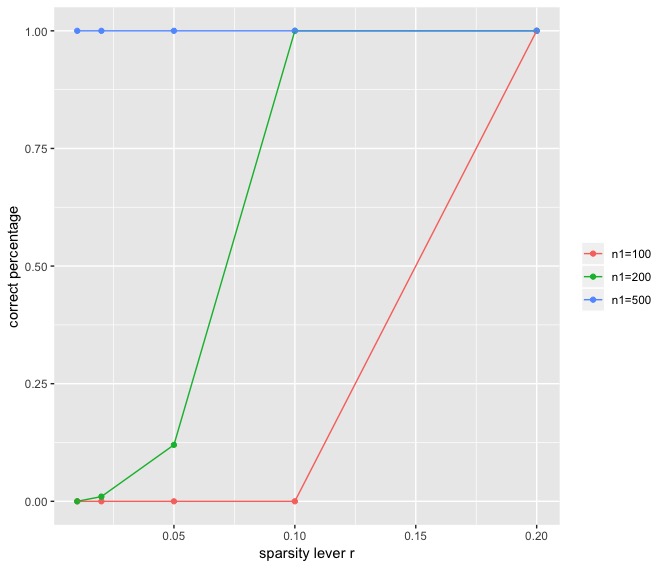}
\end{minipage}%
\begin{minipage}[t]{0.5\linewidth}
\includegraphics[width=3in]{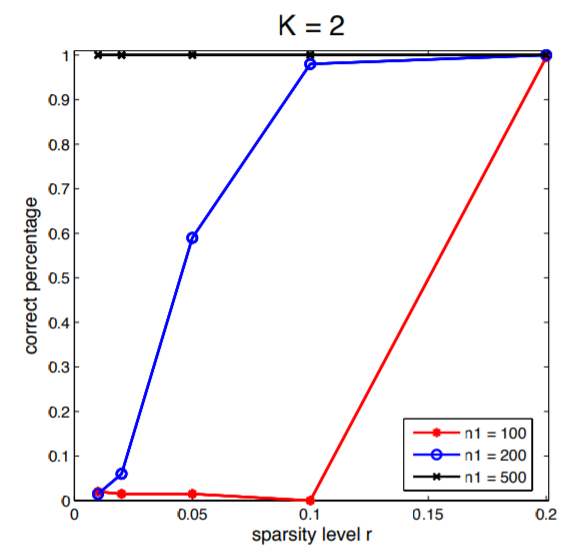}
\end{minipage}\caption{\emph{Simulation Comparison. Left: my simulation result. Right: simulaiton result from \citet{chen2018network}}}\label{fig3}
\end{figure}

\emph{Example 3.} 
I duplicate one of the simulation result in this paper. In the simulation study, the community-wise edge probability matrix $B =rB_0$, where the diagonal entries of $B_0$ are 3 and off-diagonal entries are 1. $n = 1000, K=2$. The sparsity levels are
chosen at $r \in \{0.01, 0.02, 0.05, 0.1, 0.2\}$.  Size of the first community is set to be $n_1$ and size of second community is $n-n_1$. Figure \ref{fig3} is the simulation result  from 200 simulated data sets.

We can see that there is slight difference between my simulation result and that from the paper. The main difference is that when $n_1=200$ and sparsity level is $0.05$, I only observed a correct percentage of 0.12 but the author observed the 
The difference might come from the different choice of clustring method in Step 2 in the algorithm.

The author also gave theoretical properties for their NCV under SBM. They firstly introduce two notions of community recovery consistency. One is exactly consistent recovery. Given a sequence of SBMs with $K$ blocks parameterized by $\left(g^{(n)}, B^{(n)}\right)$, we call a community recovery method $\hat{g}$ exactly consistent if $P(\hat{g}(A, K))=\left.g^{(n)}\right) \rightarrow 1$ where $A$ is a realization of SBM $\left(g^{(n)}, B^{(n)}\right)$ and the equality is up to a possible label permutation. The other  is called approximately consistent recovery. For a sequence of SBMs with $K$ blocks parameterized by $\left(g^{(n)}, B^{(n)}\right)$ and a sequence $\eta_n= o(1)$, we say $\hat{g}$ is approximately consistent with rate $\eta_n$ if,
$$\lim _{t \rightarrow \infty} P\left[\operatorname{Ham}(\hat{g}(A, K), g) \geq \eta_{n} n\right]=0,$$
where Ham($\hat{g}, g$) is the smallest Hamming distance between $\hat{g}$
and $g$ among all possible label permutations.

NCV is proved to be approximately consistent with rate $(n\rho_n)^{-1}$, with $\rho_n n/ log(n) > C$ for some constant $C$. See Theorem 2 in \citet{chen2018network}. Therefore it also has  community recovery consistency property.

\section{Conclusion}
In this paper, we first examine the statistical approach proposed by \citet{bickel2016hypothesis} for hypothesis testing under the null hypothesis $H_0: K=1$. They derived the asymptotic null distribution of the largest eigenvalue of a suitably scaled and centered adjacency matrix. However, this method exhibits a slow convergence rate, necessitating bootstrap corrections in practical applications. Subsequently, we explore the SCORE method introduced by \citet{jin2015fast}, which is designed for clustering on the Degree-Corrected Block Model (DCBM). This method effectively mitigates the impact of degree heterogeneity by utilizing the coordinate ratio of the eigenvectors of the adjacency matrix.Additionally, \citet{chen2018network} introduced the Network Cross-Validation (NCV) method to automate the selection of $K$. This method demonstrates robust performance on both the Stochastic Block Model (SBM) and DCBM, offering exact consistency. However, its primary drawback is its time-intensive nature, a challenge common to all methods based on cross-validation. Optimizing this aspect could involve identifying potential patterns in the distribution of each element or the entry-wise ratio of the eigenvectors, which might streamline the process.

\bibliographystyle{apalike}
\bibliography{bibliography}

\end{document}